\documentclass[aps,prd,twocolumn,showpacs,superscriptaddress]{revtex4}
\usepackage{graphics}
\usepackage{graphicx}
\usepackage{pstricks}
\usepackage{epsfig}

\begin{document}
\title{Flavor changing neutral currents with a fourth family of quarks}

\author{Johana A. Herrera}
\affiliation{Instituto de F\'\i sica, Universidad de Antioquia,
A.A. 1226, Medell\'\i n, Colombia.}

\author{Richard H. Benavides}
\affiliation{Instituto de F\'\i sica, Universidad de Antioquia,
A.A. 1226, Medell\'\i n, Colombia.}

\author{William A. Ponce}
\affiliation{Instituto de F\'\i sica, Universidad de Antioquia,
A.A. 1226, Medell\'\i n, Colombia.}


\begin{abstract}
{For a model with a fourth family of quarks, new sources of flavor changing neutral currents are identified by confronting the unitary $4\times 4$ quark mixing matrix with the experimental measured values of the familiar $3\times 3$ quark mixing matrix. By imposing as experimental constraints the known bounds for the flavor changing neutral currents, the largest mixing of the known quarks with the fourth family ones is established. The predictions are: a value for $|V_{tb}|$ significantly different from unity, large rates for rare top decays as $t\rightarrow c\gamma$ and $t\rightarrow cZ$, the last one reachable at the Large Hadron Collider, and large rates for rare strange decays $s\rightarrow d\gamma$ and $s\rightarrow dg$, where $g$ stands for the gluon field.}
\end{abstract}


\maketitle

\section{\label{sec:sec1}Introduction}
The so-called flavor problem encloses two of the most intriguing puzzles in modern particle physics, which are the number of fermion families in nature and the pattern of fermion masses and mixing angles, both in the quark and lepton sectors. The standard model (SM), based on the local gauge group $SU(3)_c\otimes SU(2)_L\otimes U(1)_Y$~\cite{jf}, fails to throw some insight into these two subjects. With each family being anomaly-free by itself, the SM renders, on theoretical grounds, the number of generations completely unrestricted, except for the indirect bound imposed by the asymptotic freedom of strong interactions, based on the local gauge group $SU(3)_c$, also known as quantum cromo dynamics (QCD).

Many attempts to answer the question of hierarchical quark mass matrices and mixing angles for three families have been reported in the literature, using the top quark as the only heavy quark at the weak scale \cite{hf}. Further insight into the flavor problem can be gained by contemplating the existence of additional heavy quarks.

In this analysis we study the quark mass spectrum and its mixing matrix, for a model which includes four up-type quarks and four down-type quarks, coming either from a heavy fourth family or from something else (up and down extra quark fields are present in many extensions of the SM, as, for example, in 3-3-1 models without exotic electric charges ~\cite{vl}, in $E_6$ grand unified theories ~\cite{fg}, in littlest Higgs models ~\cite{na},etc.).

With only three generations, the quark mixing matrix, called in the literature the Cabibbo-Kobayashi-Maskawa (CKM) mixing matrix ~\cite{nc}, is a $3\times 3$ unitary matrix. This unitary for models with only one SM Higgs doublet implies, first, the absence of flavor changing neutral currents (FCNC) at tree level and second, the suppression of the same FCNC at the one-loop level, due to the presence of the Glashow-Iliopoulos-Maini (GIM) mechanism~\cite{sg}.

For a model with an extra up-type quark $t^\prime$ and another extra down-type quark $b^\prime$, the quark mixing matrix becomes a unitary $4\times 4$ matrix, for which any $3\times 3$ submatrix loses its unitary character as long as the new quarks mix with the ordinary ones. One outstanding consequence of a $3\times 3$ nonunitary mixing matrix for the known quarks is the existence of new FCNC processes. Our aim in this analysis is to see how large the mixing between the ordinary quarks and exotic ones can be, in a model with four up-type quarks and four down-type quarks, without violating current experimental measurements, both in the $3\times 3$ quark mixing matrix and in the existing bounds for FCNC processes.

To gain predictability in our analysis, let us assume that the two new quarks are members of a fourth family in a trivial extension of the SM, without any other extra ingredient added (SM4). In this way, new sources of FCNC, coming either from the scalar sector or from the existence of new gauge bosons, are avoided.

The paper is organized as follows: in Sec.~\ref{sec:sec2} we review the main features of the SM  with three families and its trivial extension to four families, in Sec.~\ref{sec:sec3} we present the most general quark mass matrices for four families which is the basis of the numerical analysis carried 
through in Sec.~\ref{sec:sec4}. Finally, in Sec.~\ref{sec:sec5} we present our conclusions.

\section{\label{sec:sec2}FEATURES OF THE STANDARD MODEL} 
A brief summary of the SM is the following.
\subsection{\label{sec:sec21}Main features}
The main ingredients of the successful SM are ~\cite{jf}:
\begin{itemize}
\item A local gauge group $SU(3)_c\otimes SU(2)_L\otimes U(1)_Y$ with the flavor sector $SU(2)_L\otimes U(1)_Y$ hidden and the $SU(3)_c$ color sector confined.

\item The fermion structure of the  model with the left-handed fields belonging to doublets of $SU(2)_L$ and the right-handed fields placed in singlets, with the following particle content:
\begin{center}
$ Q_{iL}=(u_i,d_i)_L \sim (3,2,1/3),$\\
$ L_{iL}=(\nu_i,l_i^-)_L \sim (1,2,-1),$\\
$u_{iL}^c\sim(3^*,1,-4/3),$\\
$d_{iL}^c\sim (3^*,1,2/3),$\\
$l_i^c\sim (1,1,2),$
\end{center}
where the numbers in parentheses stand for $[SU(3)_c,SU(2)_L,U(1)_Y]$ quantum numbers, and $i=1,2,...,$ is a family index. Usually, $i=1,2,3$ is assumed (SM3).
\item The Higgs mechanism which triggers the spontaneous breaking of the symmetry in the flavor sector, for which the self-interacting isodoublet scalar field $\phi=(\phi^+,\phi^0)\sim (1,2,1)$, with a minimum of the scalar potential given by the vacuum expectation value $\langle \phi \rangle=(0,v/\sqrt{2})$, plays a crucial role in the theory.
\item The existence of an unbroken electric charge generator, given by
\[Q=T_{3L}+Y/2\]
associated with the massless photon field $A^\mu$, where $T_{3L}$ is the diagonal generator of $SU(2)_L$.
\end{itemize}

Some remarks about the features enunciated above are:

\begin{itemize}
\item The fermion field spectrum of the SM does not include right-handed neutrinos.
\item The electroweak precision measurements done at LEP experiments imply that the number of light neutrinos $\nu_{iL}$ is equal to three~\cite{WM}, with the three neutrinos in the flavor basis being $\nu_{1L}=\nu_{eL}$, $\nu_{2L}=\nu_{\mu L}$, and $\nu_{3L}=\nu_{\tau L}$, the neutrinos associated with the electron, muon and tauon respectively. 
\item The minimal ingredients enunciated above are not able to explain the experimental result of neutrino oscillations ~\cite{RN}, so the model must be enlarged in some way.
\item To date, there is not direct experimental evidence for the existence of the Higgs scalar field $\phi$.
\item The model is renormalizable ~\cite{jf}, with the anomalies cancelled family by family.
\item $v\approx 246 $ GeV is the electroweak scale established for the model.
\end{itemize}
Since the number of light neutrinos is just three, most people assume the existence of only three families of quarks and leptons, with the quark fields in the flavor basis for the three families being $(u_1,d_1)=(u,d)$, $(u_2,d_2)=(c,s)$, and $(u_3,d_3)=(t,b)$.

\subsection{\label{sec:sec22}SM with four families}
Determining the number of fermion families is a key goal of the upcoming experiments at the LHC~\cite{holdom}, and further at the ILC~\cite{ciftci}. This is due to the fact that the uncertainties on the measured CKM matrix elements~\cite{WM} left an open door for more quarks, with a fourth family $(t',b')$ and their mixing with the other three, not ruled out yet. Experiments at the Tevatron have already constrained the masses of a fourth family of quarks to be $m_{t^\prime}>258$ GeV and 
$m_{b^\prime}>268$ GeV~\cite{aatonen}.

Even if the existence of a fourth SM family is still an open possibility, special attention must be paid due to the necessity of including in the fermion spectrum, the lepton sector together with the quark sector, in order to cancel the anomalies and render the model theoretically consistent. Besides, data from LEP-1 established three families of fermions with light neutrinos~\cite{WM}, which however does not exclude the existence of heavy neutrinos $(m_{\nu^\prime_\tau}>M_Z/2)$.

Constraints on the masses of the fourth family fermions $t^\prime,\; b^\prime,\; \tau^\prime$, and $\nu_\tau^\prime,$ are obtained from their contributions to the electroweak corrections parameters S and T~\cite{gates}, with the one-loop contribution assuming masses sufficiently above $M_Z$. Remarkably, four family fermions with masses about 550 GeV would couple strongly to the Goldstone bosons of the electroweak symmetry breaking~\cite{holdom}, joining in this way the issue of the flavor problem with the until now, obscure spontaneous symmetry breaking mechanism.

It is clear thus, that there are not experimental or phenomenological evidence which excludes the existence of a fourth family with a heavy neutrino. Indeed, the recent electroweak precision data are equally consistent with the presence of three or four families~\cite{jhe}, whereas the four family scenario is favored if the Higgs mass is heavier than 200 GeV~\cite{kribs}.

\section{\label{sec:sec3}The quark mass spectrum.}
In ``SMN", the Standard Model for N families and just one Higgs scalar doublet $\phi$, the quark Yukawa Lagrangian can be written as:

\begin{equation}\label{lagma}
\mathcal{L}=\sum_{i=1}^N Q_{iL}^T [ \sum_{j=1}^N h_{ij}^u\phi C u_{jL}^c +  h_{ij}^d\tilde{\phi}Cd_{jL}^c]+ c.c,
\end{equation}
where $C$ is the charge conjugation operator, $\tilde{\phi}=i\tau_2\phi^*$ with $\tau_2$ an SU(2) generator, and $h_{ij}^a$, $a=u,d$ are Yukawa coupling constants.

In order to set the notation, let us write the quark mass matrices produced by the Lagrangian (\ref{lagma}) for four families. For the up quark sector and in the basis $(u,c,t,t')$, it is
\begin{equation}\label{maup}
M_U=\frac{v}{\sqrt{2}}\left(\begin{array}{cccc} 
h_{11}^u & h_{12}^u & h_{13}^u & h_{14}^u \\ 
h_{21}^u & h_{22}^u & h_{23}^u & h_{24}^u \\
h_{31}^{u} & h_{32}^{u} & h_{33}^u & h_{34}^{u} \\
h_{41}^u & h_{42}^u & h_{43}^u & h_{44}^{u} \\ 
\end{array}\right),
\end{equation}
and for the down quark sector and in the basis $(d,s,b,b')$, it is
\begin{equation}\label{madown}
M_D=\frac{v}{\sqrt{2}}\left(\begin{array}{cccc} 
h_{11}^d & h_{12}^d & h_{13}^d & h_{14}^d \\ 
h_{21}^d & h_{22}^d & h_{23}^d & h_{24}^d \\
h_{31}^d & h_{32}^d & h_{33}^d & h_{34}^d \\
h_{41}^d & h_{42}^d & h_{43}^d & h_{44}^d \\ 
\end{array}\right).
\end{equation}
$M_U$ and $M_D$ in (\ref{maup}) and (\ref{madown}) must be diagonalized in order to get the mass eigenstates, defining in this way a unitary $4\times 4$ quark mixing matrix of the form 
\begin{equation}\label{dmix}
V_{mix}\equiv V_L^uV_L^{d\dag}=\left(\begin{array}{cccc} 
V_{ud} & V_{us} & V_{ub} & V_{ub^\prime} \\ 
V_{cd} & V_{cs} & V_{cb} & V_{cb^\prime} \\
V_{td} & V_{ts} & V_{tb} & V_{tb^\prime} \\
V_{t^\prime d} & V_{t^\prime s} & V_{t^\prime b} & V_{t^\prime b^\prime}
\end{array}\right),
\end{equation}
where $V_L^u$ and $V_L^d$ are unitary $4\times 4$ matrices which diagonalize $M_UM_U^\dag$ and $M_DM_D^\dag$ respectively. $V_{mix}$ in (\ref{dmix}) defines the couplings of the physical quark states with the charged current associated with the weak gauge boson $W^+$.

\section{\label{sec:sec4}NUMERICAL ANALYSIS}
In this section we are going to see how large the Yukawa coupling constants $h_{i4}^u$, $h_{4i}^u$, $h_{i4}^d$ and $h_{4i}^d$, $i=1,2,3,4$ can be, without violating current experimental limits.

Two kinds of experimental constrains will be considered: the measured values of the $3\times 3$ quark mixing matrix and current bounds for FCNC processes.

\subsection{\label{sec:sec41}The $3\times 3$ mixing matrix}
The masses and mixing of quarks in the SM come from Yukawa interaction terms with the Higgs condensate, as can be seen from Eq.(\ref{lagma}). For $N=3$, the results are two $3\times 3$ mass matrices for the up and down quark sectors, the upper $3\times 3$ left-handed corners of matrices (\ref{maup}) and (\ref{madown}) respectively, that must be diagonalized in order to identify the mass eigenstates. The unitary quark mixing matrix, called now the CKM mixing matrix ($V_{CKM}\equiv V^u_{3L}V^{d\dag}_{3L}$) couples the six physical quarks to the charged weak current as before, where $V_{3L}^u$ and $V_{3L}^d$ are now the diagonalizing unitary $3\times 3$ matrices.

The matrix $V_{CKM}$ has been parametrized in the literature in several different ways, but the most important fact related with this matrix is that most of its entries have been measured with high accuracy, with the following experimental results~\cite{see}:

\begin{widetext}
\begin{equation}\label{maexp}
 V_{exp}=
\left(\begin{array}{ccc}
0$.$970\leq  V_{ud}\leq 0$.$976 & 0$.$223\leq V_{us}\leq 0$.$228 & 0$.$003\leq V_{ub}\leq0$.$005\\
0$.$219\leq V_{cd}\leq 0$.$241 & 0$.$90\leq V_{cs}\leq 1$.$0 & 0$.$039 \leq V_{cb}\leq 0$.$040\\
0$.$006\leq V_{td}\leq 0$.$008 & 0$.$036\leq V_{ts}\leq 0$.$044 & V_{tb} \geq0$.$78
\end{array}\right).
\end{equation}
\end{widetext}

The numbers quoted in matrix (\ref{maexp}) are conservative, in the sense that they are related to the direct experimental measured values with the largest uncertainties taken into account, without bounding the numbers to the orthonormal  constrains on the rows and columns of $V_{CKM}$. In this way, we leave the largest room for possible new physics, respecting the measured values in $V_{exp}$.

\subsection{\label{sec:sec42}FCNC}
The unitary character of $V_{CKM}$ implies flavor diagonal couplings of all the neutral bosons of the SM (such as Z boson, Higgs boson, gluons and photon) to a pair of quarks, giving as a consequence that no FCNC are present at tree level. At one-loop level, the charged currents generate FCNC transitions via penguin and box diagrams~\cite{jf}, but they are highly suppressed by the GIM mechanism~\cite{sg}. For example, FCNC processes in the charm  sector $(c\rightarrow u\gamma)$ were calculated in the context of the SM in Ref.~\cite{GB}, giving a branching ratio suppressed by 15 orders of magnitude, leaving in this way a large window of opportunities for new physics in charm decays.

To date, the following FCNC branching bounds have been established in several experiments:

\begin{itemize}
 \item $\mathcal{B}r[s\rightarrow d\gamma(dl^+l^-)]<10^{-8}$ ~\cite{sp}
\item $\mathcal{B}r[c\rightarrow ul^+l^-]<4\times 10^{-6}$ ~\cite{vm}
\item $\mathcal{B}r[b\rightarrow s\gamma,d\gamma (\gamma\longrightarrow l^+l^-)]<5\times 10^{-7}$ ~\cite{VM},
\end{itemize}
With $l=e,\mu$.

\subsection{\label{sec:sec43}Textures}
In order to explain the known quark masses and mixing angles, several ansatz for up and down mass matrices have been suggested in the literature~\cite{hf}, some of them including the so-called texture zeros~\cite{SW}. In particular, symmetric mass matrices with four and five texture zeros were studied in detail in Refs.~\cite{HF} and~\cite{LI}, respectively. Unfortunately, precision measurements of several entries in the mixing matrix rule out most of the suggested simple structures.

As far as the mixing matrix is concerned, our numerical analysis found the following six texture zeros symmetric mass matrices, quite appropriate

\begin{equation}\label{uptex}
M_u =\frac{h_tv}{\sqrt{2}}\left(\begin{array}{ccc} 
0 & 7\lambda^6 & 0 \\ 
7\lambda^6 & 0 & 4\lambda^2  \\
0 & 4 \lambda^2  & 4
\end{array}\right),
\end{equation}

\begin{equation}\label{dowtex}
M_d =\frac{h_bv}{\sqrt{2}}\left(\begin{array}{ccc} 
0 & 3\lambda^6 & 0 \\ 
3\lambda^6 & 4\lambda^5 & 0  \\
0 &  0  & 2\lambda^2 
\end{array}\right), 
\end{equation}
where $h_t$ and $h_b$ are Yukawa coupling constants fixed by the top and bottom quark masses, respectively.

The former ansatz for up and down quark mass matrices resembles the Georgi-Jarlskog conjecture~\cite{HG}, with the extra ingredient of being compatible with a new kind of flavor symmetry and its perturbative breaking as proposed by Froggatt and Nielsen~\cite{CD}, including a second order effect at the level of the bottom quark mass, implied by the entry $(M_d)_{33}\sim 2\lambda^2$.

As we will see next, a value of $\lambda\approx 0.21$ in matrices (\ref{uptex}) and (\ref{dowtex}) is able to reproduce all the experimental constrains quoted in matrix (\ref{maexp}).

\subsection{\label{sec:sec44}The $4\times 4$ mixing matrix}
In this section we are going to analyze the mixing matrix $V_{mix}$ for a model with four up-type quarks and four down-type quarks (as in a fourth family model, for example), and confront it with the experimental values quoted in $V_{exp}$ in matrix (\ref{maexp}). For this purpose, the ansatz suggested by $M_u$ in matrix (\ref{uptex}) for the up quark sector, and by $M_d$ in matrix (\ref{dowtex}) for the down quark sector, are going to be used for the upper-left $3\times 3$ mass submatrices in (\ref{maup}) and (\ref{madown}) respectively.

To check the validity of our approach, let us start by using zero entries for all the non diagonal Yukawa coupling constants in the fourth row and fourth column of $M_U$ and in the fourth row and fourth column of $M_D$, using for the diagonal entries the numerical values $h_{44}^u=10h_t$ and $h_{44}^d=10h_b$, which imply masses for the exotic quarks $t^\prime$ and $b^\prime$ at the TeV scale. Then, using the value $\lambda=0.21$  we diagonalize numerically $M_U$ and $M_D$ and then calculate $V_{mix}^0=V_L^uV_L^{d\dag}$, obtaining, up to three decimal places, the result

\begin{equation}\label{mix0}
V_{mix}^{(0)} =\left(\begin{array}{cccc} 
0.974 & 0.227 & -0.003 & 0 \\ 
0.227 & -0.973 & 0.044 & 0  \\
-0.007 &  0.040  & 0.999 & 0 \\
0 & 0 & 0 & 1 
\end{array}\right),
\end{equation}
where the negative values are just a consequence of the unitary character of $V_{mix}^{(0)}$ (they can be changed to positive values by a redefinition of the quark fields).

The absolute values  of all the nonzero entries in $V_{mix}^{(0)}$ agree fairly well with the experimental values quoted in matrix (\ref{maexp}). The zeros in the fourth row and fourth column of $V^{(0)}_{mix}$, coming from the fact that there are not mixing between ordinary and the fourth family  quarks at this level, imply the absence of new FCNC effects coming from the mixing matrix at this order zero approach.

Notice the values $h_{44}^u=10h_t$ and $h_{44}^d=10h_b$, introduced in order to cope with the Tevatron experimental limits~\cite{aatonen} for $m_{t^\prime}$ and $m_{b^\prime}$, which imply Yukawa coupling constants of 2.5 (as normally expected for a fourth family). These large Yukawa coupling constants for the fourth family are just on the limit of the perturbative regime, suggesting the existence of new physics at the TeV scale. These strong couplings allow one to speculate on the possibility of a heavy quark condensate, able to break in a consistent way the electroweak symmetry, as presented, for example, in Refs.~\cite{holdom, hill}.

The numerical analysis which follows aims to set upper bounds on the fourth rows and fourth columns of $M_U$ and $M_D$, using as phenomenology constrains the values of the matrix $V_{exp}$ in (\ref{maexp}). Entries in $M_U$ and $M_D$ of order 1-10 will imply a strong mixing of the ordinary quarks with the exotic ones; entries of the order of $\lambda$, $\lambda^2$, and $ \lambda^3$ will imply weak mixing; and entries of the order of $\lambda^4$,$\lambda^5$, and $\lambda^6$ will imply very weak mixing. 

To continue the analysis, notice next that the constrains on the fourth row of $M_U$ and on the fourth row of $M_D$ coming from the matrix $V_{exp}$ in (\ref{maexp}) are a second order effect because those two rows refer to the mixing of the left-handed fourth family quark components  with the right-handed quark components of the ordinary ones, with $V_{exp}$ related only to the mixing of the left-handed components as it is explicit in the definition of $V_{mix}$. In order to gain predictability in our analysis, we are going to assume first a left-right symmetry in our model, which in turns implies symmetric $4\times 4$ mass matrices as generalizations of the symmetric ones $M_u$ and $M_d$ in (\ref{uptex}) and (\ref{dowtex}).

The systematic random numerical analysis using MATHEMATICA subroutines, throws as a result that the maximum mixing allow, without violating the experimental bounds of the mixing matrix (5), or the known bounds for FCNC, are given by the following set of numbers [with $(M_U)_{44}$ and $(M_D)_{44}=$ taken as before]:

\begin{center}
$(M_U)_{41}=(M_U)_{14}=h_tv\lambda^4/\sqrt{2},$\\
$(M_U)_{42}=(M_U)_{24}=h_tv\lambda^3/\sqrt{2},$\\
$(M_U)_{43}=(M_U)_{34}=h_tv/\sqrt{2},$\\
$(M_D)_{41}=(M_D)_{14}=h_bv\lambda^6/\sqrt{2},$\\
$(M_D)_{42}=(M_D)_{24}=h_bv\lambda^4/\sqrt{2},$\\
$(M_D)_{43}=(M_D)_{34}=10h_bv/\sqrt{2};$
\end{center}

\noindent
which produce the following unitary mixing matrix:

\begin{widetext}
\begin{equation}\label{mixud}
V_{mix}^{ud} =\left(\begin{array}{cccc} 
0.974 & -0.227 & 0.003 & -2.0\times 10^{-3} \\ 
0.227 & 0.973 & -0.040 & 2.1\times 10^{-2} \\
0.007 &  0.044  & 0.920 & -0.39 \\
-3.7\times 10^{-4} & -3.6\times 10^{-3} & 0.39 & 0.92 
\end{array}\right).
\end{equation}
\end{widetext}

\begin{figure}
\includegraphics{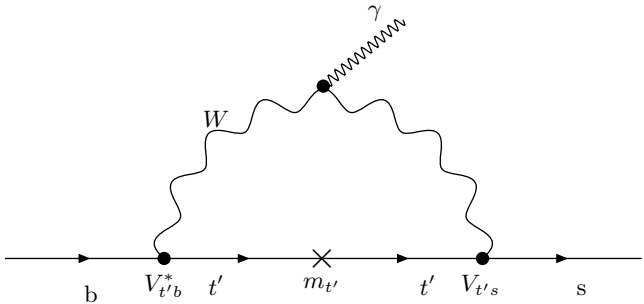}
\caption{\label{fig1}One loop diagram contributing to the FCNC $b\longrightarrow s\gamma$}
\end{figure}

\subsection{\label{sec:sec45}New FCNC processes}
With the numbers in (\ref{mixud}) let us evaluate first the branching ratio ${\cal B}r(b\rightarrow s\gamma)$ using Fig.~(\ref{fig1}), calculated with the expectator model, scaled to the semileptonic decay $b\rightarrow q_il\nu_l,\; q_i=c,u$, and without including QCD corrections (which are small for the $b$ sector~\cite{jf}). According to Eq.(6) in the first paper in Ref. \cite{GB}, this branching ratio  (where all the entries are easily understood) is given by
\begin{equation}\label{fincal}
{\cal B}r(b\rightarrow s\gamma)\approx\frac{3\alpha}{2\pi}
\frac{|V_{t^\prime b}^*V_{t^\prime s}F^Q(x)|^2}{[f(x_c)|V_{cb}|^2+f(x_u)|V_{ub}|^2]}B_{B\rightarrow Xl\nu_l},
\end{equation}
where $\alpha = 1/137$ is the fine structure constant, $B_{B\rightarrow Xl\nu_l}\approx 0.1$ is the branching ratio for semileptonic $b$ meson decays taken from Ref.~\cite{WM}, $x=(m_{t^\prime}/M_W)^2$, $x_c=m_c/m_b$, and $x_u=m_u/m_b$, $F^Q(x)$ is the contribution of the internal heavy quark line to the electromagnetic penguin given by
\begin{eqnarray*}
F^Q(x)&=&Q\left[\frac{x^3-5x^2-2x}{4(x-1)^3}+\frac{3x^2\ln{x}}{2(x-1)^4}\right] \\
&+&\frac{2x^3+5x^2-x}{4(x-1)^3}-\frac{3x^3\ln{x}}{2(x-1)^4},
\end{eqnarray*}
where $Q=2/3$ for $t^\prime$ in the quark propagator [$Q=-1/3$ and $x=(m_{b^\prime}/M_W)^2$ when $b^\prime$ propagates] and $f(x_i)$ is the usual phase space factor in semileptonic meson decay, given by~\cite{jf}
\[f(x)=1-8x^2+8x^6-x^8-24x^4\ln{x}.\]

The numerical evaluations for $m_t^\prime=1.0$ TeV, $m_c=1.5$ GeV, $m_b=4.9$ GeV, and $m_u=2.0$ MeV produce the following values: $F^{2/3}(x)\approx 0.6325$, $f(x_c)\approx 0.5$, and $f(x_u)\approx 1$. Plug in these values in Eq.~(\ref{fincal}), and using the numbers obtained for $V_{mix}^{ud}$ in Eq.~(\ref{mixud}) for the couplings of the physical quark states, gives
\[{\cal B}r(b\rightarrow s\gamma)\approx 3.4\times 10^{-7},\]
close to the experimental measured bound~\cite{VM}.

A similar analysis shows that 
\[{\cal B}r(b\rightarrow d\gamma)=\frac{|V_{t^\prime d}|^2}{|V_{t^\prime s}|^2}{\cal B}r(b\rightarrow s\gamma)\approx 3.6\times 10^{-9},\]
which is in agreement with the bound quoted in Sec.~(\ref{sec:sec42}). 

In a similar way we can evaluate ${\cal B}r(s\rightarrow d\gamma)$ scaled to the semileptonic decay $s\rightarrow ul\nu_l$, which is given now by 
\begin{equation}\label{fincs}
{\cal B}r(s\rightarrow d\gamma)\approx\frac{3\alpha}{2\pi}
\frac{|V_{t^\prime s}^*V_{t^\prime d}F^{2/3}(x)|^2}{f(x_u)|V_{us}|^2}B_{K\rightarrow \pi l\nu_l}.
\end{equation}
With $B_{K\rightarrow \pi l\nu_l}\approx 5\times 10^{-2}$ taken from Ref.~\cite{WM}, we get 
\[{\cal B}r(s\rightarrow d\gamma)\approx 2.4\times 10^{-15},\]
also in good agreement with the experimental bound quoted in Sec.~(\ref{sec:sec42}).

Now let us evaluate ${\cal B}r(c\rightarrow u\gamma)$ scaled to the semileptonic decay $c\rightarrow q_jl\nu_l$, where $q_j=s,d$. The branching ratio is
\begin{equation}\label{finccu}
{\cal B}r(c\rightarrow u\gamma)\approx\frac{3\alpha}{2\pi}
\frac{|V_{cb^\prime}^*V_{ub^\prime}F^{-1/3}(x)|^2}{[f(x_s)|V_{cs}|^2+f(x_d)|V_{cd}|^2]}B_{D\rightarrow X_s l\nu_l},
\end{equation}
where $x_s=m_s/m_c,\; x_d=m_d/m_c$. With $B_{D\rightarrow X_s l\nu_l}\approx 0.2$ taken from Ref.~\cite{WM}, $F^{-1/3}(x)\approx 0.3856$, $f(x_s)\approx 0.97$ for $m_s=150$ MeV and $f(x_d)\approx 1$, for $m_d=5$ MeV, we get 
\[{\cal B}r(c\rightarrow u\gamma)\approx 1.89 \times 10^{-13},\]
2 orders of magnitude larger than the SM prediction~\cite{GB}, but still unobservably small.

We proceed our study of FCNC for ordinary quarks, with the top quark sector. The new FCNC ${\cal B}r(t\rightarrow c\gamma)$ and ${\cal B}r(t\rightarrow u\gamma)$ predicted for the top quark in the context of a model with an extra up-type quark and one extra down-type quark, scaled to the semileptonic decay $t\rightarrow q_k l\nu_l,\; q_k=b,s,d$ are given by 
\begin{equation}\label{finct}
\frac{{\cal B}r(t\rightarrow c\gamma)}{B_{T\rightarrow Xl\nu_l}}\approx\frac{3\alpha}{2\pi}
\frac{|V_{tb^\prime}^*V_{cb^\prime}F^{-1/3}(x)|^2}{[f(x_b)|V_{bt}|^2+f(x_s)|V_{st}|^2+f(x_d)|V_{dt}|^2]}
\end{equation}
which implies 
\[{\cal B}r(t\rightarrow c\gamma)\approx 0.5\times 10^{-7}B_{T\rightarrow Xl\nu_l},\]
which is large as far as the semileptonic branching ratio $B_{T\rightarrow Xl\nu_l}$ measured for the top quark gets comparatively large, and much larger than $10^{-14}$, the SM prediction~\cite{saav}.

Finally we find 
\[{\cal B}r(t\rightarrow u\gamma)=\frac{|V_{ub^\prime}|^2}{|V_{cb^\prime}|^2}{\cal B}r(t\rightarrow c\gamma)\approx 0.5\times 10^{-9} B_{T\rightarrow Xl\nu_l}.\]

\subsection{\label{sec:sec46}FCNC processes for the fourth family}
As can be seen from the former calculations, the GIM cancellation which occurs in the chiral $U(4)\times U(4)$ limit, does not proceed  now because the branching ratios are proportional to $F^Q(x)^2$, which is a function of $(x=m_{q^\prime}/M_W)^2\gg 1$, for $q^\prime=t^\prime,\; b^\prime$.

To make predictions for the fourth family a hierarchy between the heavy quarks must be assumed; for example, for $m_{t^\prime}>m_{b^\prime}>m_t$, and scaling the branching ratio to the semileptonic decay $b^\prime\rightarrow Ul\nu_l$ for $U=t,c,u$, we get
\begin{equation}\label{finbpb}
\frac{{\cal B}r(b^\prime\rightarrow b\gamma)}{B_{B^\prime\rightarrow X_Ul\nu_l}}\approx\frac{3\alpha}{2\pi}
\frac{|V_{t^\prime b^\prime}^* V_{t^\prime b}F^{2/3}(x)|^2}{[f(x_t)|V_{tb^\prime}|^2+f(x_c)|V_{cb^\prime}|^2+f(x_u)|V_{ub^\prime}|^2]},
\end{equation}
which for $m_t=173$ GeV produces the result 

\[{\cal B}r(b^\prime\rightarrow b\gamma)\approx 1.5\times 10^{-3} B_{B^\prime\rightarrow X_Ul\nu_l}, \]
a value large enough to be detected at the LHC, even if the branching ratio $B_{B^\prime\rightarrow X_Ul\nu_l}$ is of the order of $10^{-2}$.

Similar numerical results follow for the branching ratio $t^\prime\rightarrow t\gamma$ for the hierarchy $m_t<m_{t^\prime}<m_{b^\prime}$.

\section {\label{sec:sec5} Conclusions}
The basic motivation of the present work was to study the up and down quark mass matrices and their mixing, in the context of a model with an extra up-type quark and one extra down-type quark, allowing for maximal mixing between ordinary and exotic quarks without violating current experimental constrains in the quark mixing matrix and in bounds coming from FCNC processes.

Just to be mentioned, another result from our random numerical analysis is the fact that the mixing in the down sector could be as large as $(M_D)_{43}=(M_D)_{34}=100h_dv$, which produces a value  $V_{tb}\sim 0.88$, but with a larger branching ratio ${\cal B}r(b\rightarrow s\gamma)$ than the bound quoted in Ref.~\cite{VM}.

With the era of the LHC approaching, copious production of top quarks is expected. With the genuine heavy quark physics just beginning, there is serious  hope that FCNC processes can reach the $10^{-5}$ or $10^{-6}$ sensitivity at LHC, or other future $e^-e^+$ colliders. We also expect a well measured $|V_{tb}|$ value at the first LHC run.

Notice that a rare top decay as for example $t\rightarrow cZ$ should be, in the context of our analysis, of the order of 
\[{\cal B}r(t\rightarrow cZ)=\frac{4\pi}{\sin^2(2\theta_W)}{\cal B}r(t\rightarrow c\gamma)\approx 
10{\cal B}r(t\rightarrow c\gamma),\]
a value not far from the LHC capability.

From our study, the main implications of the possible existence of one extra up-type and one extra down-type quark are:
\begin{itemize}
\item Very large FCNC for the new heavy quarks, to be measured at the LHC.
\item A large branching ratio ${\cal B}r(b\rightarrow s\gamma)$.
\item A value $0.90\leq V_{tb}\leq 0.94$ which largely violates the ordinary $3\times 3$ unitary condition of the known mixing matrix.
\item Rare top decay as for example $t\rightarrow cZ$ to be detected in the near future.
\end{itemize}
The former results hold in a much larger class of extensions to the SM which include heavy quarks, beyond the trivial four family extension studied here.

Branching for FCNC decays in the top quark sector are small, not because the matrix elements are small, but because the semileptonic top quark decays are enhanced by the same matrix elements.

Notice also for example that processes like $s\rightarrow dg$, where $g$ stands for a gluon field, are dominated by the gluon penguin which is proportional to $\alpha_s$(1 GeV)$\approx 0.1$, 1 order of magnitude larger than $\alpha$, the fine structure constant present for the electromagnetic penguin. So, this process should be sensitive to the present experiments at the B factories.

The main result obtained from this analysis is the large violation of the GIM mechanism produced by the mere existence of heavy quark flavors which mix with the ordinary ones, violating the unitary condition of the ordinary $3\times 3$ mixing matrix, as it should be expected from general grounds.

To conclude this analysis let us say that larger values for FCNC processes than the ones calculated above can be obtained for almost all the channels studied, just by allowing mixing only in one of the two quark sectors.

\section*{Acknowledgments}
We thank Enrico Nardi for a critical reading of the original manuscript.


\begin{thebibliography}{}
\bibitem[1]{jf}
For an excellent compendium of the SM see: J. F. Donoghue, E. Golowich and B. Holstein, \textit{Dynamics of the Standard Model}, (Cambridge University Press, Cambridge, England, 1992).

\bibitem[2]{hf}
For a review with extensive references, see: H.Fritzsch and Z.Z.Xing, Prog. Part Nucl.Phys.\textbf{45}, 1 (2000).

\bibitem[3]{vl}
J.C.Montero, F.Pisano and V.Pleitez, Phys. Rev. D\textbf{47}, 2918 (1993);  
R. Foot, H.N. Long and T.A. Tran, Phys. Rev. D \textbf{50}, R34 (1994). H.N. Long, Phys. Rev. D \textbf{53}, 437 (1996); \textit{ibid} \textbf{54}, 4691 (1996); V. Pleitez, Phys. Rev. D \textbf{53}, 514 (1996);D.A.Guti\'errez,W.A.Ponce and L.A.S\'anchez,Eur,Phys. J.C\textbf{46}, 497 (2006); D.Chang and H.N.Long,Phys. Rev. D\textbf{73}, 053006 (2006).

\bibitem[4]{fg}
F.G\"{u}rsey, P.Ramond and P.Sikiev, Phys. Lett. B\textbf{60} 177 (1976); S.Okubo, Phys. Rev. D\textbf{16}, 3528 (1977);J.L.Hewett and T.G.Rizzo, Phys. Rep. \textbf{183}, 193 (1989).

\bibitem[5]{na}
N.Arkani-Hamed, A.G.Cohen and Georgi, Phys. Lett. B\textbf{513}, 232 (2001). For a review and complete references see: M.Schmaltz and D,Tucker-Smith, Ann.Rev.Nucl.Part.Sci. \textbf{55}, 229 (2005) 

\bibitem[6]{nc}
N.Cabbibo, Phys, Rev. Lett. \textbf{10}, 531 (1963);M.Kobayashi and T. Maskawa, Prog. Theor. Phys. \textbf{49}, 652 (1973).

\bibitem[7]{sg}
S.L.Glashow, J.Iliopoulos, and L.Maini, Phys. Rev, D\textbf{2}, 1285 (1970).

\bibitem[8]{WM}
W.M.Yao \textit{et al}, Journal of Physics G\textbf{33}, 1 (2006); http://pdglive.lbl.gov/. 

\bibitem[9]{RN}
For a recent review, see R.N.Mohapatra and Y.Smirnov, Annu. Rev. Nucl. Part. Sci. \textbf{56}, 569 (2006), and references therein.

\bibitem[10]{holdom}
B.Holdom, JHEP \textbf{0608}: 076 (2006); JHEP \textbf{0703}: 063 (2007); ``{\it The heavy quark search at the LHC}" arXiv:0705.1736v1.

\bibitem[11]{ciftci}
A.K. Ciftci, R.Ciftci and S.Sultansoy, Phys. Rev. D\textbf{72}, 053006 (2005).

\bibitem[12]{aatonen}
T. Aatonen {\it et al} (CDF collaboration), Phys. Rev. D\textbf{76}, 072006 (2007).

\bibitem[13]{gates}
E. Gates and J. Terning, Phys. Rev. Lett.\textbf{67}, 1840 (1991); T. Appelquist and J. Terning, Phys. Lett. B\textbf{315}, 139 (1993).

\bibitem[14]{jhe}
H.J.He, N.Polonsky and S.F.Su, Phys. Rev. D\textbf{64}, 053004 (2001); V.A. Navikov, L.B.Okun, A.N. Rozonov and M.I. Vysotsky, Phys. Lett. B\textbf{529}, 111 (2002).

\bibitem[15]{kribs}
G.D.Kribs, T.Plehn, M.Spannowsky and T.M.P. Tait, Phys. Rev. D\textbf{76}, 075016 (2007).

\bibitem[16]{see}
See section 11 and pages from 867 to 881 in Ref.[8] and references therein. See also the WEB pages: ``http://www.utfit.org" and ``http://ckmfitter.in2p3.fr/".

\bibitem[17]{GB}
G.Burdman,E.Golowich,J.Hewett and S. Pakvasa, Phys. Rev. D \textbf{52}, 6383 (1995); \textit{ibid}, D \textbf{66}, 014009 (2002).

\bibitem[18]{sp}
See page 43 in Ref.[8].

\bibitem[19]{vm}
V.M.Abazov \textit{et al}, Phys. Rev. Lett \textbf{100}, 101801 (2008).

\bibitem[20]{VM}
V.M.Abazov \textit{et al}, Phys. Rev. Lett \textbf{94}, 071802 (2005);\textit{ibid} Phys. Rev. D \textbf{76}, 092001 (2007).

\bibitem[21]{SW}
S.Weinberg, in ``A Festschhrift for I.I.Rabi'', Trans. N.Y. Acad. Sci. Ser. II, Vol.\textbf{38}, 185 (1977); F.WilczeK and  A.Zee, Phys. Lett. B\textbf{70}, 418 (1977); Phys. Rev. Lett. \textbf{42}, 421 (1979); J.Chakrabarti, Phys. Rev. D\textbf{20}, 2411 (1979).

\bibitem[22]{HF}
H.Fritzsch and Z.Z.Xing, Phys. Lett. B\textbf{555}, 63 (2003) and references therein.

\bibitem[23]{LI}
Luis Iba\~nez and G.G.Ross, Phys. Lett. B\textbf{332}, 100 (1994).

\bibitem[24]{HG}
H.Georgi and C.Jarlskog, Phys. Lett. B\textbf{86}, 297 (1979).

\bibitem[25]{CD} 
C.D.Froggatt and H.B.Nielsen, Nucl. Phys. B\textbf{147}, 277 (1979).

\bibitem[26]{hill}
W.A.Bardeen C.T.Hill and M.Lindner, Phys. Rev. D\textbf{41}, 1647 (1990); R.S.Chivukula \textit{et al}, Phys. Rev. D\textbf{59}, 075003 (1999).

\bibitem[27]{saav}
J.A.Aguilar Saavedra, Acta Phys. Polon. B\textbf{35}, 2695 (2004).
\end{thebibliography}
\end{document}